# Internal or external magma oceans in the earliest protoplanets −
# perspectives from nitrogen and carbon fractionation

Damanveer S. Grewal[1,2]*, Johnny D. Seales[1], Rajdeep Dasgupta[1]

[1]Department of Earth, Environmental, and Planetary Sciences, Rice University, 6100 Main Street, MS 126, Houston, TX 77005, USA

[2]Division of Geological and Planetary Sciences, California Institute of Technology, 1200 E California Blvd, Pasadena, CA 91125, USA

*correspondence: dgrewal@caltech.edu

**Abstract**

        Protoplanets growing within ~1 Ma of the Solar System's formation underwent large-scale melting due to heat released by the decay of [26]Al. When the extent of protoplanetary melting approached magma ocean (MO)-like conditions, alloy melts efficiently segregated from the silicates to form metallic cores. The nature of the MO of a differentiating protoplanet, i.e., internal or external MO (IMO or EMO), not only determines the abundances of life-essential volatiles like nitrogen (N) and carbon (C) in its core and mantle reservoirs but also the timing and mechanism of volatile loss. Whether the earliest formed protoplanets had IMOs or EMOs is, however, poorly understood. Here we model equilibrium N and C partitioning between alloy and silicate melts in the absence (IMO) or presence (EMO) of vapor degassed atmospheres. Bulk N and C inventories of the protoplanets during core formation are constrained for IMOs and EMOs by comparing the predicted N and C abundances in the alloy melts from both scenarios with N and C concentrations in the parent cores of magmatic iron meteorites. Our results show that in comparison to EMOs, protoplanets having IMOs satisfy N and C contents of the parent cores with substantially lower amounts of bulk N and C present in the parent body during core formation. As the required bulk N and C contents for IMOs and EMOs are in the sub-chondritic and chondritic range, respectively, N and C fractionation models alone cannot be used to distinguish the prevalence of these two end-member differentiation regimes. A comparison of N and C abundances in chondrites with their peak metamorphic temperatures suggests that protoplanetary interiors could lose a substantial portion of their N and C inventories with increasing degrees of thermal metamorphism. Provided the thermal metamorphism induced-loss of N and C from the protoplanetary interiors prior to the onset of core formation was efficient, the earliest formed protoplanets, as predicted by previous thermo-chemical models, are more likely to have undergone IMO differentiation resulting in the formation of N- and C-poor cores and mantles overlain by N- and C-rich undifferentiated crusts.

## 1. Introduction

        The physical and chemical make-up of protoplanets that formed within ~1 Ma of the formation of the Solar System dictates the chemical compositions of present-day rocky planets (Elkins-Tanton et al., 2011; Grewal et al., 2021b; Hirschmann et al., 2021). Investigations of iron meteorites provide key insights into the formation and evolution of these earliest formed protoplanets that underwent at least one large-scale melting event (Goldstein et al., 2009). The accretion ages of the parent bodies of a few magmatic irons (originating from cores which were once fully molten) are almost contemporaneous with that of Calcium-Aluminum-rich Inclusions (CAIs – the oldest solids which date 'time zero' in Solar System history) (Kruijer et al., 2017). Heat released by the decay of [26]Al (t[1/2] ~ 0.7 Myr) was the primary cause of melting and differentiation of their interiors (Hevey and Sanders, 2006; Sahijpal et al., 2007).

Thermochemical models predict that protoplanets with radii greater than ~20 km accreting within ~1 Ma after CAIs approached MO-like conditions (i.e., surpassed ~50 vol.% silicate melting) before undergoing complete metal-silicate separation (Kaminski et al., 2020; Neumann et al., 2012). This mode of differentiation is in contrast with incomplete/complete metal-silicate separation in unmolten or partially molten silicates (e.g., non-magmatic irons and achondrites likes ureilites, acapulcoites-lodranites, winonaites, and brachinites (McSween, 1989)) of parent bodies which accreted rather late, were of smaller sizes, and/or experienced an early loss of [26]Al by explosive volcanism (Neumann et al., 2012; Sugiura and Fujiya, 2014). In this study, we focus on the earliest accreted protoplanets with relatively large sizes which likely underwent complete metal-silicate separation in largely molten silicates.

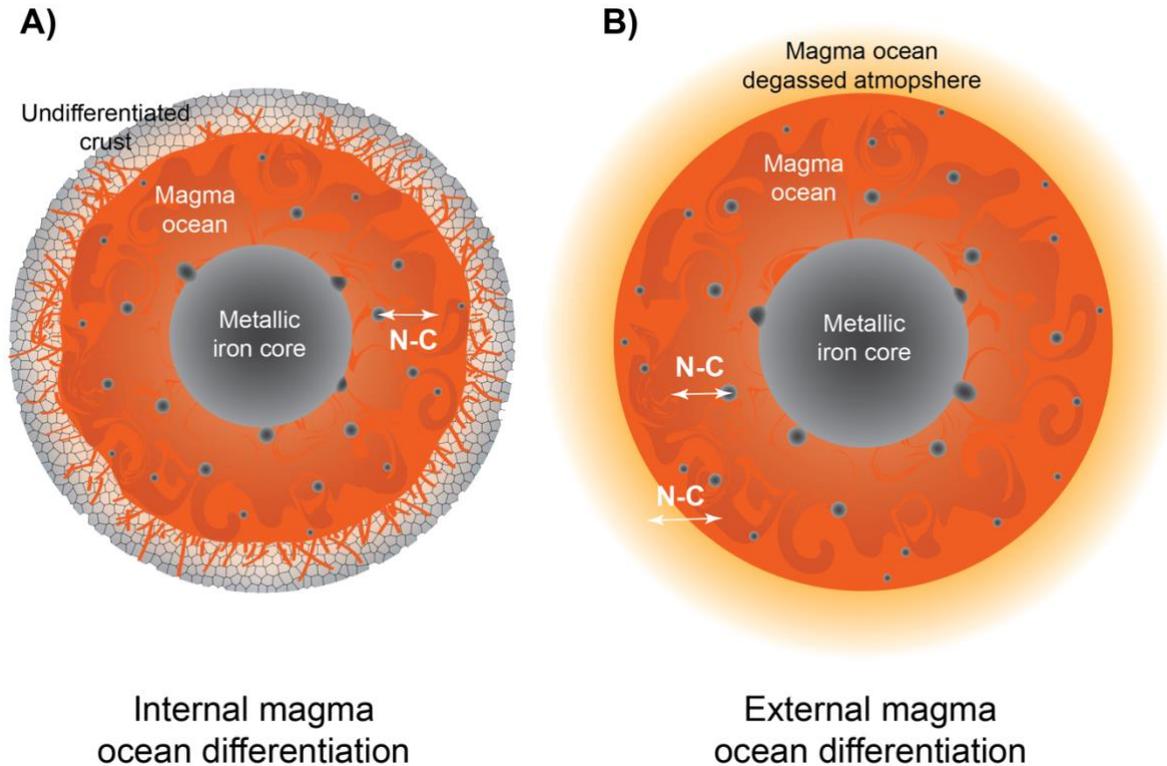

***Figure 1:*** *Illustrations depicting **A**) internal and **B**) external magma ocean differentiation regimes. **A**) Fractionation of N and C between alloy melt and silicate melt in an IMO. **B**) Fractionation of N and C between atmosphere, silicate melt, and alloy melt in an EMO. Figure modified from Elkins-Tanton et al. (2011).*

An interior outward melting in these protoplanets resulted in the formation of either internal or external MOs (IMOs or EMOs) – depending on whether MOs were overlain by unmolten shells or not (Elkins-Tanton et al., 2011; Grewal et al., 2021b; Hirschmann et al., 2021) (Fig. 1). EMOs and IMOs are alternately postulated to explain several aspects of the meteorite record. For instance, severe depletion of moderately volatile elements (MVEs; e.g., Cr and Cl) and highly volatile elements (HVEs; e.g., C and N) coupled with their isotopic fractionation trends in howardite-eucrite-diogenites (HEDs) and angrites was explained via vapor-silicate melt exchange at the surface of EMOs (e.g., Abernethy et al., 2013; Zhu et al., 2019). Similarly, atmosphere-MO-core exchange in EMOs was posited to explain N depletion in the silicate reservoirs of rocky protoplanets (Grewal et al., 2021b) as well as the low C/S ratios in the parent cores of magmatic iron meteorites (Hirschmann et al., 2021). Vapor-silicate melt exchange at the surface of MOs was also used to explain the heavy isotope signatures of refractory elements (e.g., Si and Mg) in HEDs and angrites relative to chondrites (Hin et al., 2017; Young et al., 2019). On the other hand, the strongest evidence for IMOs emanates from the combined paleomagnetic data for CV, CM, and Rumuruti chondrite parent bodies as well as IIE iron meteorite parent body (IMPB hereafter) where chondritic shells were underlain by convective MOs and molten cores (e.g., Carporzen et al., 2011; Maurel et al., 2020). Numerical models simulating the thermal evolution of protoplanetary interiors heated by the decay of $^{26}$Al also call for MO differentiation underneath solid, conductive shells (Elkins-Tanton et al., 2011; Hevey and Sanders, 2006; Kaminski et al., 2020; Neumann et al., 2012; Sahijpal et al., 2007).

Whether rocky protoplanets had IMOs or EMOs during their differentiation has important implications for HVE abundances in the resulting core and mantle reservoirs. In an IMO, elemental abundances in the metallic and silicate reservoirs are set by alloy melt-silicate melt equilibration (Fig.1A). Whereas an EMO interacts with the overlying atmosphere such that vapor-silicate melt exchange at its surface sets up elemental abundances in the MO (provided the rate of atmospheric loss is slow enough to allow vapor-silicate melt equilibration). This exchange subsequently determines the proportion of an element partitioned into the core during alloy melt-silicate melt equilibration (Fig.1B). In other words, elemental abundances in the resulting core and mantle reservoirs of a protoplanet undergoing EMO differentiation are set by coupled exchange between vapor, silicate melt, and alloy melt (Grewal et al., 2021b; Hirschmann et al., 2021). The re-distribution of highly volatile elements like nitrogen (N) and carbon (C), which efficiently fractionate between vapor, silicate melt, and alloy melt, is especially sensitive to these end-member MO differentiation regimes. In addition to having common origins in primitive organic matter, N and C show comparable chemical characteristics during protoplanetary and planetary processing (e.g., Alexander et al., 1998, 2007; Grewal et al., 2019b; Sephton et al., 2003). Importantly, there is sufficient data on the thermodynamic relationships of these elements, i.e., their vapor pressure induced solubilities in surficial silicate melts as well as their alloy melt-silicate melt partition coefficients at low to moderate pressures, to constrain their partitioning between vapor, silicate melt, and alloy melt (e.g., Dasgupta and Grewal, 2019; Grewal et al., 2019a; Libourel et al., 2003; Ni and Keppler, 2013; Tsuno et al., 2018; Yoshioka et al., 2019).

Comparisons of the predictions of N and C abundances in MOs and cores (based on atmosphere-MO-core and MO-core equilibration models for EMOs and IMOs, respectively) with their estimated abundances in primitive silicate or alloy melts (based on the meteoritic record) can be used to compare the relative prevalence of these end-member differentiation regimes. Iron meteorites provide direct estimates of N and C abundances in protoplanetary cores (Grewal et al., 2021c; Hirschmann et al., 2021). Therefore, the cores of IMPBs which did not suffer extensive evaporation related losses post-disruption of their parent bodies can provide reliable estimates on N and C abundances in the protoplanetary cores. Using iron meteorites to compare the relative prevalence of IMOs and EMOs in their earliest formed protoplanets provides additional advantages: 1) Grouped irons sample multiple parent bodies; therefore, they are statistically the largest representation of the earliest formed protoplanets in the Solar System (magmatic irons –groups IC, IIAB, IIC, IID, IIF, IIG, IIIAB, IIIE, IIIF, IVA, and IVB and non-magmatic irons – groups IAB (MG, sL) and IIE (Goldstein et al., 2009)). 2) Hf-W chronometry combined with thermal modeling predicts the accretion of magmatic IMPBs ~0.3-1 Ma after CAI formation – thereby recording the compositions of the earliest formed protoplanets in the Solar System (Kruijer et al., 2017). 3) Based on the nucleosynthetic anomalies of Mo, it has been determined that IMPBs record the growth of protoplanets both in the inner and the outer regions of the solar protoplanetary disk – which makes them a more generalized archive of protoplanetary growth (groups IAB (MG, sL), IC, IIAB, IIE, IIIAB, IIIE, and IVA belonging to the non-carbonaceous (NC) cluster purportedly had an inner Solar System origin, and groups IIC, IID, IIF, IIIF, and IVB belonging to the carbonaceous (CC) cluster had an outer Solar System origin (Kruijer et al., 2017)).

Because inner and outer Solar System protoplanets have different physical and chemical characteristics based on the differences in the chemical compositions of the accreting material in different regions of the disk (Worsham et al., 2019), iron meteorites potentially capture variable modes of N and C incorporation in the metallic cores of their parent bodies. Consequently, in this study we use N and C abundances in magmatic iron meteorites in conjunction with atmosphere-MO-core and MO-core equilibration partitioning models of N and C to study the MO environments of the earliest formed protoplanets.

## 2. Nitrogen and carbon contents in the parent cores of iron meteorites

Carbon is present in iron meteorites not only as a dissolved component in kamacite, taenite, and plessite, but also in accessory phases like graphite, cohenite ((Fe, Ni, Co)$_3$C), and haxonite ((Fe, Ni)$_{23}$C$_6$) in some groups of iron meteorites (Goldstein et al., 2017). Accounting, or lack thereof, of C in the accessory phases results in the discrepancies between the reported C abundances in iron meteorites (Goldstein et al., 2017; Lewis and Moore, 1971). As samples within a given group of magmatic iron meteorites record different stages of fractional crystallization (Goldstein et al., 2009), their C contents may not be representative of C abundances of their parent cores if there was a significant partitioning of C between solid and liquid metal during cooling of the core. The effect of fractional crystallization on the partitioning of an element between solid and liquid metal in magmatic irons is generally constrained by comparing its abundances with Ni and Ir contents of similar samples within a given group – early crystallizing alloys are Ni-poor/Ir-rich (Goldstein et al., 2009). Unlike other trace elements like Ge, Ga, As, etc., C does not show any correlations with indices of fractional crystallization in magmatic iron groups (Goldstein et al., 2017). Therefore, individual samples for a given group roughly capture the C concentrations in their parent core. Hirschmann et al. (2021) reported upper bounds for C contents of the parent cores of several groups of iron meteorites by matching the activity of C in solid Fe-Ni-C metal with that of the equilibrating Fe-Ni-C-S liquid (Table 1). Combining all previous data with the thermodynamic models, Hirschmann et al. (2021) estimated the representative C contents of the parent cores of magmatic iron meteorite groups IC, IIAB, IIC, IID, and IIIAB to be within the range of 100-1100 ppm. We have used this range of C contents to constrain our numerical models. Carbon contents of the extremely volatile depleted IVA and IVB groups – likely a result of vapor-metal melt exchange at the surface of stripped metallic cores post-disruption of their parent bodies (Yang et al., 2010, 2008) – were not included in our model calculations.

Unlike C, almost all N (up to ~98%) in iron meteorites resides in kamacite and taenite (Prombo and Clayton, 1993). To constrain the effect of fractional crystallization, we plotted the N contents against Ir contents for samples within a given group (e.g., IIIAB and IVA groups in Supplementary Fig. 1). N and Ir contents in both IIIAB and IVA groups are uncorrelated. The cause behind the observed scatter in the data (analogous to the observations of Goldstein et al. (2017) in C vs Ni plots for IIIAB and IVA groups) requires further investigation but simply could be a result of rapid diffusion of N in the solid metal post crystallization of the core. Importantly, the lack of correlation in N vs Ir plots suggests that the iron meteorites do not record any fractionation of N between solid and liquid metal during cooling of the parent cores. Therefore, similar to C, the N contents in the parent cores for a given group of magmatic iron meteorites can also be directly inferred from the samples of that group without placing them in the fractional crystallization sequence. Average N contents are highest in IAB and IC groups and lowest in IVA and IVB groups (Table 1). IIE group also shows a similar N depletion as IVA and IVB groups. As a result, IIE, IVA and IVB irons likely do not record the primitive N contents of their parent cores. Therefore, we did not use N contents of these anomalously volatile-depleted groups as constraints for our models. Based on the interquartile range of other magmatic irons, i.e., IC, IIAB, IIC, IID, IIF, and IIIAB groups, we used 5-50 ppm N as a representative range for the N contents of the parent cores.

*Table 1: Parameters relevant for iron meteorite parent bodies and other protoplanets.*

| Parent body | [a]Measured N (ppm) | [b]Estimated C (ppm) | Core-mantle mass ratio | Core mass fraction | Radius (km) | S in core (wt.%) | $T$ (°C) | log$fO_2$ (IW) |
|---|---|---|---|---|---|---|---|---|
| IAB (MG) | 15-64 | - | - | - | - | - | - | −2.5 to −1.5 |
| IC | 21-62 | 260 | - | - | - | - | - | - |
| IIAB | 8-17 | 170 | - | - | - | 17 | 1325 | - |
| IIC | 4-11 | 230 | 0.20 | 0.17 | - | - | - | - |
| IID | 24-32 | 190-1100 | - | - | - | 7 | 1445 | - |
| IIE | 0.4-2 | - | - | - | - | - | - | −2.5 to −1.5 |
| IIF | 16 | - | 0.11-0.28 | 0.1-0.22 | - | - | - | - |
| IIIAB | 15-36 | 64-230 | - | - | - | 12 | 1400 | - |
| IVA | 0.5-5 | 4-610 | 0.20 | 0.17 | 483 | 6 | 1445 | - |
| IVB | 0.6-1 | 11-210 | 0.09 | 0.08 | 277 | 0 | 1615 | −1 |
| [c]SBT | - | 87 | 0.11 | 0.1 | - | - | - | - |
| Vesta | - | - | - | 0.05-0.21 | 260 | - | - | −2.4 to −2.2 |
| [d]APB | - | - | - | 0.07-0.29 | >260 | - | - | −1.9 to −0.9 |

[a]Interquartile range of measured N contents in iron meteorites. Data source: Compilation in Grewal et al. (2021b)
[b]Estimated C contents in parent cores of iron meteorites. Data source: Hirschmann et al. (2021)
[c]SBT = South Bryon Trio
[d]APB = Angrite Parent Body

### 3. Contribution of nebular ingassing and dissolution into core forming alloy melts towards nitrogen and carbon contents in iron meteorites

The primary goal of this study is to decipher the amount of N and C that was partitioned into protoplanetary cores based on the two end-member regimes of MO differentiation. Therefore, it is critical to differentiate between the amount of N and C present in the primordial metal (formed during nebular condensation) prior to the accretion of rocky bodies and the amount of N and C incorporated into the metallic melts exclusively during core formation. Thermodynamic calculations predict that equilibration of nebular gas with the condensing γ-Fe metal can ingas ~0.01 ppm (Grewal et al., 2021c) and ~1 ppm C (Lewis et al., 1979) (Fig. 2). These values are distinctly lower than N and C contents of even the most volatile-depleted irons. Hence, nebular ingassing into the primordial metal did not contribute significantly to the eventual N and C budgets of protoplanetary cores. On the other hand, melting and segregation of alloy melts in the interiors of protoplanets takes place at significantly higher pressures (~$10^4$ bar) and can potentially enrich the alloy melts with significant amounts of N and C. For instance, at pressures relevant for core-mantle differentiation in planetesimal-sized rocky bodies (Steenstra et al., 2017, 2016), the limits for N and C incorporation into molten metals – as given by N solubility in the alloy melts at vapor saturation for Ni-free alloys (Speelmanns et al., 2018) and C solubility in the alloy melts at graphite saturation (e.g., Dasgupta and Walker, 2008) – lie within the range of 0.1-5 wt.% and 5-8 wt.%, respectively (Fig. 2). This means that almost entire N and C inventory of protoplanetary cores was incorporated into the alloy melts during core formation.

Alloy melts efficiently segregate to the center of a protoplanet via interconnected networks only when the silicate matrix has undergone a significant amount of melting (Taylor, 1992). This should have also resulted in efficient elemental exchange between alloy and silicate melts. Consequently, alloy melt-silicate melt equilibration during core-mantle differentiation must have set N and C abundances in the parent cores of iron meteorites.

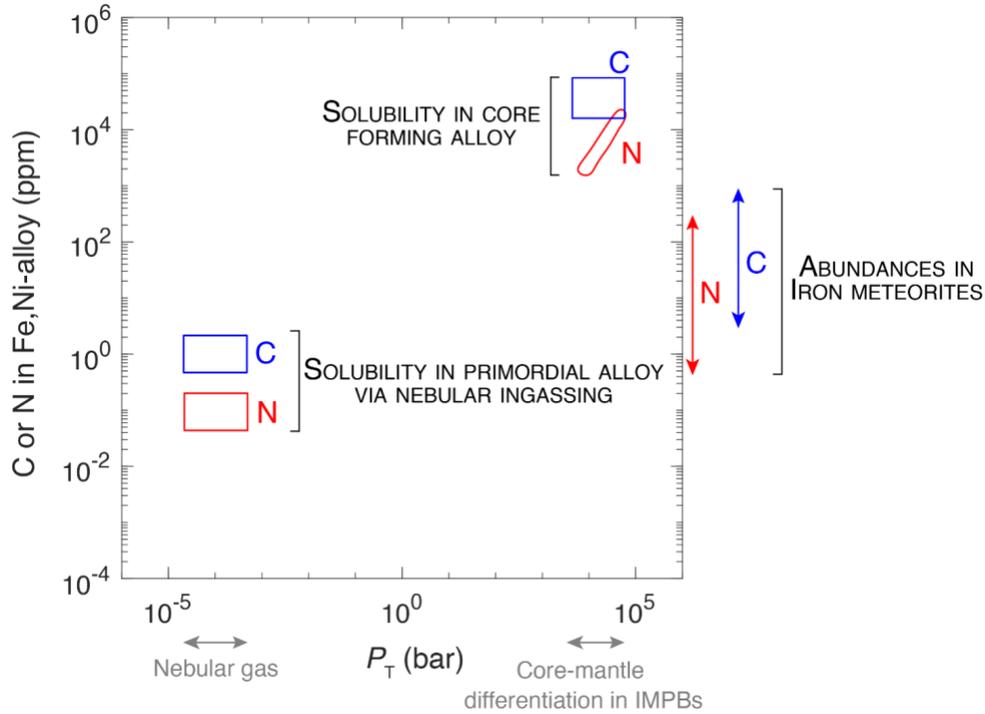

***Figure 2:*** *Nitrogen and carbon solubility in alloy melts as a function of total pressure. Due to extremely low partial pressures of N and C in the nebular gas, only ~0.1 ppm N and ~ 1 ppm C can be incorporated via nebular ingassing into the primordial alloy which condensed out of the nebular gas. These values are substantially lower than N and C contents of iron meteorites. Due to core-mantle differentiation in asteroid-sized bodies taking place at significantly higher pressures, N and C solubility in core forming alloy melts is ~1000-10,000 ppm and 20,0000-50,000 ppm, respectively. Data sources: Pressure - Nebular gas (Palme et al., 2014); Core-mantle differentiation in asteroid-sized bodies (Steenstra et al., 2017, 2016). Solubility in primordial alloy via nebular ingassing – N (Grewal et al., 2021c); C (Lewis et al., 1979). Solubility in core forming alloy melt – N (Speelmanns et al., 2018); C (Dasgupta and Walker, 2008). Abundances in iron meteorites – N (Data compilation in Grewal et al., 2021b); C (Goldstein et al., 2017).*

## 4. Incorporation of nitrogen and carbon into the protoplanetary cores via IMOs and EMOs

### 4.1 *Model setup*

In EMOs, vapor pressure induced solubility dictates N and C abundances in the silicate melts, and consequently the amount of N and C available for partitioning between alloy and silicate melts (Fig. 1B). Therefore, in EMOs the amount of N and C partitioned into protoplanetary cores was calculated based on coupled atmosphere-silicate melt-alloy melt equilibration models. Whereas in IMOs, silicate melt-alloy melt equilibration alone controls the incorporation of N and C into protoplanetary cores (Fig. 1A).

The calculations on coupled atmosphere-silicate melt-alloy melt equilibration models followed the thermodynamic framework used in several previous studies (e.g., Grewal et al., 2021a; Hirschmann et al., 2021; details are reported in the supplementary section). Nitrogen and carbon exchange between atmosphere and silicate melts was calculated based on their vapor pressure based solubilities (refer to Supplementary Section for details). Following the template of Keppler and Golabek (2019), C dissolution solely as anhydrous $CO_3^{2-}$ or CO was calculated using the Henry's law constants from Ni and Keppler (2013) ($K_{CO}$ = 0.016 ppm/MPa) and Yoshioka et al. (2019) ($K_{CO_2}$ = 0.155 ppm/MPa), respectively.

$$C_C^{MO} = K_{CO}. fCO + K_{CO_2} fCO_2 \qquad (Eq. 1)$$

where $C_C^{MO}$ represents concentration of C in the MO and $fCO$ and $fCO_2$ represent fugacities of CO and $CO_2$, respectively, in the vapor phase.

$fCO$ and $fCO_2$ were calculated based on the equilibrium reaction:

$$CO_{(g)} + \tfrac{1}{2} O_{2(g)} = CO_{2(g)} \qquad (Eq. 2)$$

where relations of equilibrium thermodynamic parameters can be calculated from:

$$\Delta G^{T,P} = G_f^{T,1\,bar}(CO_2) - G_f^{T,1\,bar}(CO) + RT \ln fCO_2 - \tfrac{1}{2} RT \ln fO_2 - RT \ln fCO = 0 \qquad (Eq. 3)$$

where $\Delta G^{T,P}$ is the change in Gibbs free energy at any $P$ and $T$ and is equal to zero at equilibrium; $G_f^{T,1\,bar}(CO_2)$ and $G_f^{T,1\,bar}(CO)$ are the standard Gibbs free energy of formation of $CO_2$ and $CO$, respectively. NIST-JANAF thermochemical tables were used to estimate the relevant values.

We used the two-species model of Libourel et al. (2003) (dissolution as $N_2$ and $N^{3-}$ in oxidized and reduced conditions, respectively (Grewal et al., 2020)) to determine N solubility in the silicate melts as a function of $pN$ and $fO_2$:

$$C_N^{MO} = 0.06\, p_N + 5.97\, p_N^{1/2}\, fO_2^{-3/4} \qquad (Eq. 4)$$

where $C_N^{MO}$ represents concentration of N in the MO and $p_N$ is the partial pressure of N in the overlying atmosphere.

Nitrogen and carbon exchange between alloy and silicate melts was calculated based on their alloy and silicate melt partition coefficients. Almost all experimental data on the partitioning of N ($D_N^{alloy/silicate}$) and C ($D_C^{alloy/silicate}$) between alloy and silicate melts till date has been obtained at high activities of N and C, i.e., alloy and/or silicate melts containing wt.% N and C. The N and C contents of protoplanetary cores are, however, 1-4 orders of magnitude lower (Grewal et al., 2021c; Hirschmann et al., 2021). This raises an important question – can the experimental $D_N^{alloy/silicate}$ and $D_C^{alloy/silicate}$ values obtained at high activities of N and C be applied to numerical models simulating alloy melt-silicate melt equilibration in systems containing ppm level

N and C? For the applicability, N and C dissolution in alloy and silicate melts must follow Henrian behavior, i.e., activity coefficients of N and C in alloy and silicate melts must not change with variations in bulk N and C contents of the system. Although the validity of Henry's law has not been tested for N-bearing systems, recent studies have examined its validity for C partitioning between alloy and silicate melts. Using thermodynamically modeled activity-composition relationships, Gaillard et al. (2022) showed that the activity coefficient of C in the alloy melts ($\gamma_C^{alloy\,melt}$) can increase by a factor of 4-5 with increasing activity of C from 0.001 (C-poor system) to 1 (graphite-saturated system). Correcting for $\gamma_C^{alloy\,melt}$, Gaillard et al. (2022) predicted that $D_C^{alloy/silicate}$ values for C-poor systems could be an order of magnitude higher relative to those determined for C-rich systems in previous experimental studies. Using high pressure ($P$)-temperature ($T$) experiments, Grewal et al. (2021a) confirmed the increase in $\gamma_C^{alloy\,melt}$ with increasing activity of C. However, contrary to the predictions of Gaillard et al. (2022), a strong effect of $\gamma_C^{alloy\,melt}$ is not directly reflected in the $D_C^{alloy/silicate}$ values of Grewal et al. (2021a). Even though the experimentally determined $D_C^{alloy/silicate}$ values of Grewal et al. (2021a) increase with increasing activity of C, those values were still well within the range of the predictions for C-rich systems. Similar observations were also made in the experimental studies of Kuwahara et al. (2021, 2019) (note that the silicate melt compositions of Kuwahara et al. (2019) were anomalously $B_2O_3$-rich due to the utility of BN capsules). Therefore, it is possible that the effect of change in the activity of C on $D_C^{alloy/silicate}$ is complex and is not captured by accounting for $\gamma_C^{alloy\,melt}$ term alone. Importantly, all the above-mentioned studies dealt with S-free systems. Whereas the S contents of the parent cores of IMPBs vary between 0 and 17 wt.% (Chabot, 2004). Aligning with the experimental data of Grewal et al. (2021a) and Kuwahara et al. (2021, 2019), which show that $D_C^{alloy/silicate}$ values for C-poor systems are within the predicted range for C-rich systems, we used the recent parametrized $D_C^{alloy/silicate}$ equation from Fischer et al. (2020) (based on experimental data for C-rich systems) as it accounts for the effect of S content in the alloy melt. We also compared the results with the parametrized $D_C^{alloy/silicate}$ equation by Grewal et al. (2021a) for C-poor systems (see Supplementary for the equations). In the absence of $D_N^{alloy/silicate}$ data at N-poor conditions and following the lead from C, we assumed Henrian behavior for N partitioning between alloy and silicate melts and used the recent parametrized $D_N^{alloy/silicate}$ equation from Grewal et al. (2021b) (see Supplementary for the equation). We would like to caution the readers that a lack of $D_N^{alloy/silicate}$ and $D_C^{alloy/silicate}$ values for N-poor and C-poor systems, respectively, in the relevant *P-T-X* space can potentially present limitations to the results of this study. Future experimental work on this front is necessary to test the validity of the model calculations and predictions of this study.

Vapor pressure-based solubilities and alloy melt-silicate melt partition coefficients depend on several independent and dependent parameters that define the physical and chemical architecture of protoplanets during atmosphere-MO-core differentiation. These include extent of alloy melt-silicate melt equilibration, core-mantle mass ratio, *P-T* of alloy melt-silicate melt equilibration, alloy melt composition, and oxygen fugacity ($fO_2$). Based on the available data for IMPBs, we varied the relevant thermodynamic parameters as follows: core-mantle mass ratio = 0.09-0.28; S content of alloy melt = 0-17 wt.%; *T* of alloy melt-silicate melt equilibration = 1325-1615 °C; $fO_2$ of core-mantle differentiation = IW–3 to IW–1; radius of parent body = 250-500 km (Fig. 3; Table 1; see supplementary section for details). Given $D_N^{alloy/silicate}$ and $D_C^{alloy/silicate}$ vary little between ambient pressure to shallow MO conditions (Dasgupta et al., 2013; Grewal et al., 2019a), *P* of alloy melt-silicate melt equilibration was fixed at 0.1 GPa (based on the predicted values for Vesta and angrite parent body). As liquidus temperatures of the parent cores are directly dependent on S content of the cores (Kruijer et al., 2014), *T* of alloy melt-silicate melt equilibration in our models was determined by S content of the alloy melt.

Bulk N and C contents involved during core-mantle differentiation were varied within 2- 2000 ppm and 50-50,000 ppm, respectively. The upper limits are based on the N and C contents of the most volatile-rich carbonaceous chondrites (e.g., Alexander et al. (2012, 2007) and lower limits are set at 0.1% of the upper limits to account for N and C loss from the parent bodies prior to the formation of alloy and silicate melts. To explore the entire parameter space that reproduces N and C contents of the primitive cores of IMPBs via IMOs and EMOs, we performed inverse Monte Carlo simulations for our thermodynamic models. Six independent parameters were used in the thermodynamic framework of the inverse Monte Carlo simulations – bulk N and C content, core-mantle mass ratio, S content of the alloy melt, $fO_2$ of core-mantle differentiation, and size of the parent body. The simulations that matched the N and C content of the primitive cores of magmatic IMPBs (5-50 ppm and 100-1100 ppm, respectively) were deemed to be successful. The values of the six independent parameters that yielded successful solutions were recorded. To yield a statistically significant solution space, the simulations were iterated to output 10,000 successful solutions.

### 4.2 *Results*

The distribution of N and C between atmosphere, silicate melts, and alloy melts is controlled by the interplay between vapor-based solubilities and alloy melt-silicate melt partition coefficients within the explored parameter space. Our simulations for EMO differentiation regime show that more than ~97% of the bulk N and C inventory in the protoplanets during core formation resides in their atmospheres (Fig. 4A, B). The cores and MOs respectively contain only ~0.1-3% and ~0.003-0.4% of the bulk N inventory (Fig. 4A, 5) and ~0.3-3% and ~0.003-0.01% of the bulk C inventory (Fig. 4B, 5). The solution space using the parametrized $D_C^{alloy/silicate}$ equation from Grewal et al. (2021a) based on data for C-poor systems suggests slightly higher bulk C values owing to lower $D_C^{alloy/silicate}$ values in C-poor systems relative to C-rich ones (Fig. 5B). For a given bulk N and C content, the total partial pressures of all N and C bearing species (*p*N and *p*C) in the atmosphere scales as a function of the gravitational constant and radius of the protoplanet. A relatively small radii of protoplanets explored in our models results in lower *p*N and *p*C ensuing in lower dissolution of N and C in MOs, limiting the amount of N and C available for fractionation between alloy and silicate melts. Because N and C are siderophile elements for core-mantle differentiation conditions applicable for IMPBs, the cores contain proportionally higher amounts of N and C relative to MOs. At a fixed bulk N and C content, the wide range of N and C abundances in the cores and MOs for both EMOs and IMOs are caused by the variations in $D_N^{alloy/silicate}$ and $D_C^{alloy/silicate}$ values within the parameter space explored for core-mantle mass ratio, S content of the alloy melt, $fO_2$ of core-mantle differentiation, and radius of the body.

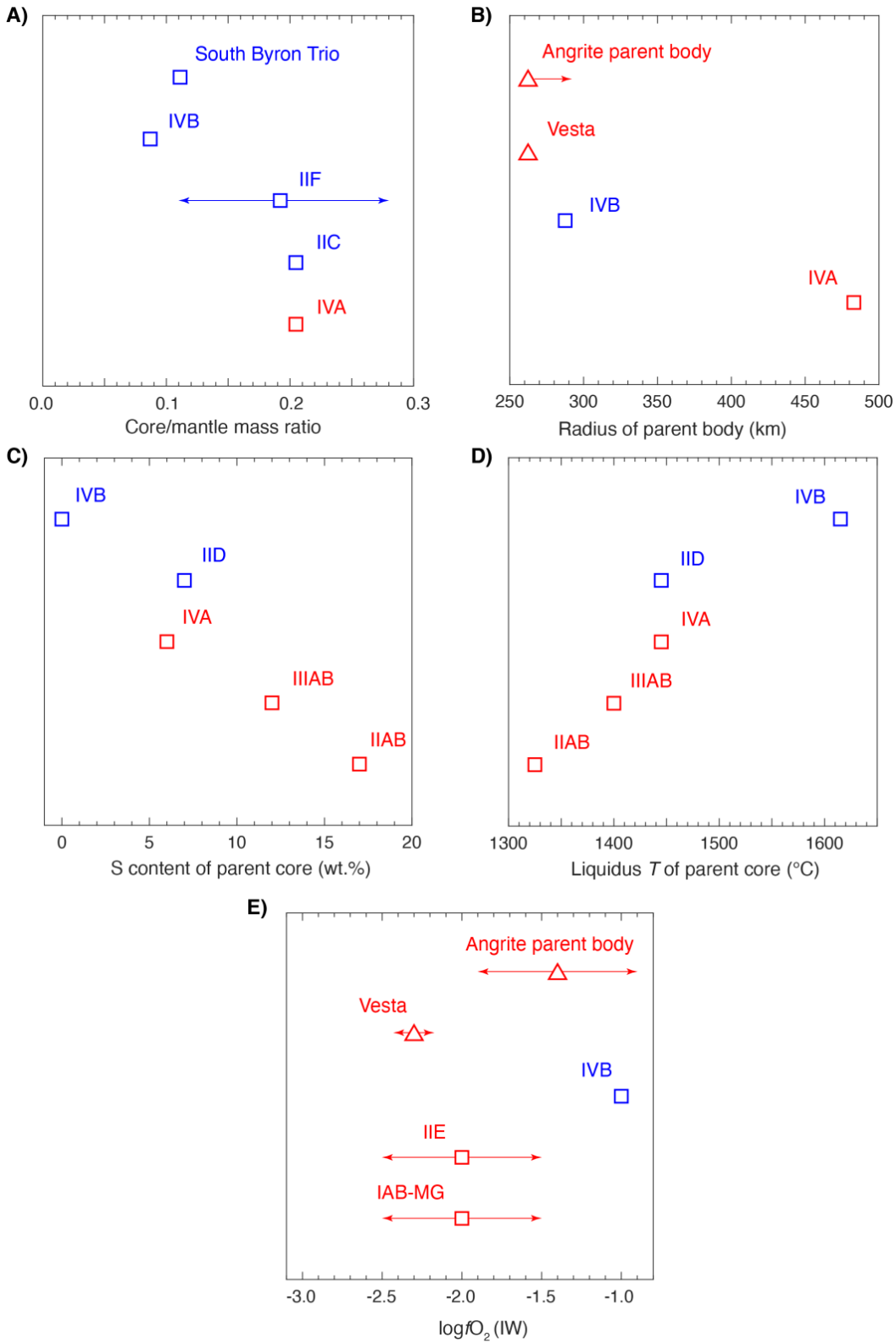

***Figure 3:*** *List of parent body parameters for different groups of iron meteorites, Vesta, and angrite parent body.* ***A****) Core-mantle mass ratio,* ***B****) radius of parent body,* ***C****) S content of parent core,* ***D****) Liquidus T of parent core, and* ***E****) fO₂ of alloy melt-silicate melt equilibration of different groups of iron meteorites. Data for Vesta and angrite parent body is supplemented in the plots only where the constraints for iron meteorite parent bodies are limited (****B****) and* ***E****)). Red and blue symbols represent iron meteorite groups belonging to the NC and CC cluster, respectively. Arrows represent the range of variation for a given parameter.*

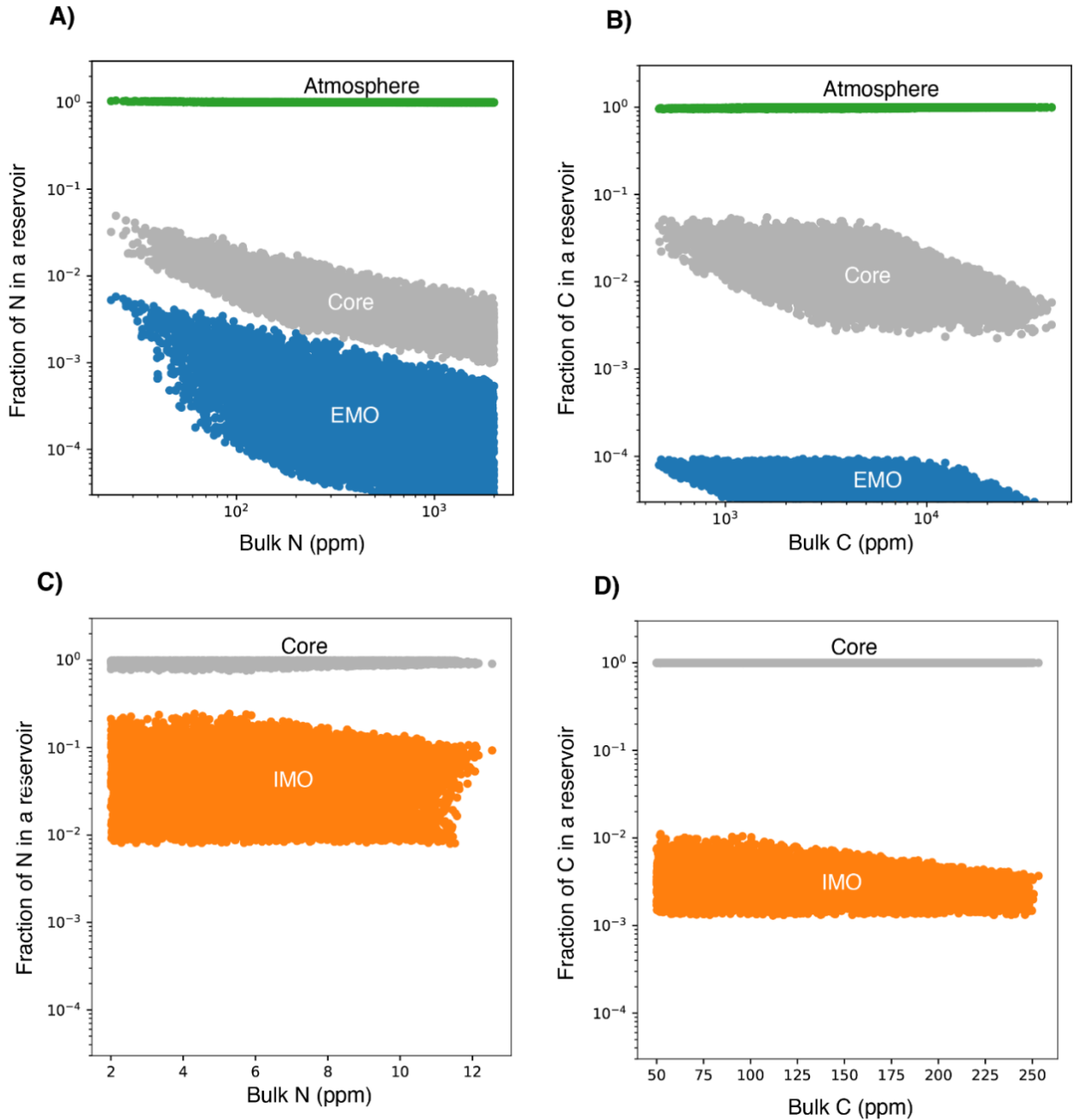

**Figure 4:** *Results of inverse Monte Carlo simulations depicting the re-distribution of N and C between constituent reservoirs of protoplanets during core formation by **A, B**) EMO and **C, D**) IMO differentiation. For EMO differentiation, almost all **A**) N and **B**) C resides in the atmosphere of protoplanets. Therefore, higher bulk N and C contents are required in an EMO relative to an IMO to explain N and C abundances in the parent cores of IMPBs. As N and C are siderophile elements for the parameter space explored for IMPBs, metallic cores contain proportionally higher N and C relative to silicate melts for both EMOs and IMOs.*

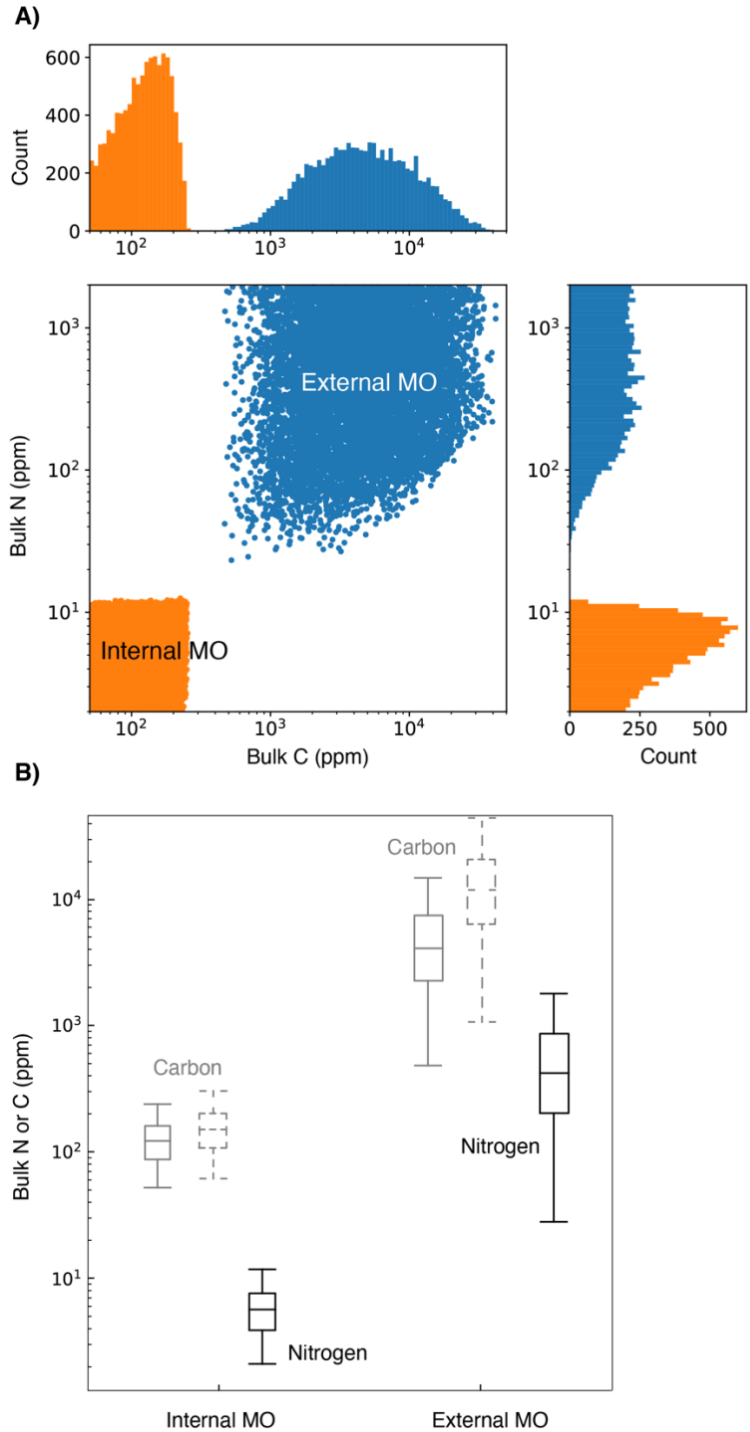

**Figure 5:** *Statistical distribution of the amounts of N and C that present in a protoplanet during core formation to explain their contents in the parent cores of magmatic IMPBs via IMO and EMO differentiation. Thermodynamic models combined with inverse Monte Carlo simulations predict that in comparison with EMOs, distinctly lower N and C contents involved in core formation can satisfy their contents in parent cores of IMOs. In **A**) the entire solution space is shown for both IMOs and EMOs. In **B**) the solution space is shown using box and whisker plots. Solid box and whisker plots for C are calculations based on the $D_C^{alloy/silicate}$ equation from Fischer et al. (2020) (based on the experimental data for C-rich systems) and dashed box and whisker plots are calculations based on the $D_C^{alloy/silicate}$ equation from Grewal et al. (2021a) (based on the experimental data for C-poor systems). Box represents the median, first and third quartile values of N and C contents and the whiskers represent the minimum and maximum values.*

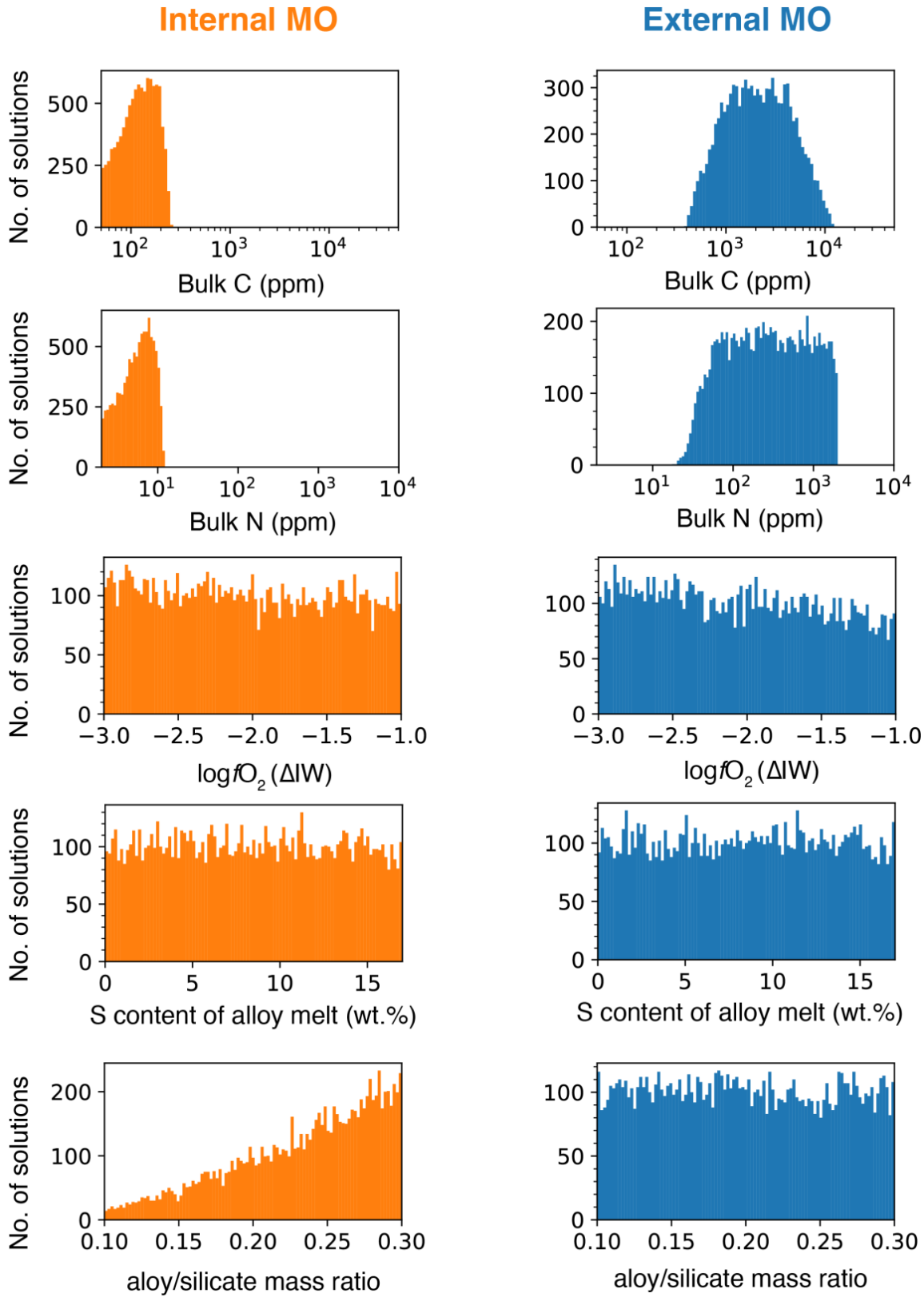

**Figure 6:** *The solution space of EMO and IMO differentiation regimes for all independent parameters. While the inverse Monte Carlo simulations yield solutions for distinctly different bulk N and C contents for EMOs and IMOs, the solution space is spread almost evenly for fO₂ of core-mantle differentiation, S content of the alloy melt, and core-mantle mass ratio.*

As the IMOs do not have overlying atmospheres, C and N are re-distributed only between the cores and MOs based on $D_N^{alloy/silicate}$ and $D_C^{alloy/silicate}$ values. The cores contain more than 80% of the bulk N inventory, and 0.8-20% of the bulk N resides in the MOs (Fig. 4C). For C, more than 99% of bulk C resides in the cores while the MOs contain 0.1-1% of the bulk C inventory (Fig. 4D). Analogous to EMOs, the solution space for $D_C^{alloy/silicate}$ equation based on C-poor data requires slightly higher bulk C contents relative to C-rich systems (Fig. 5B). A higher proportion of C than N in the cores is a direct result of $D_C^{alloy/silicate}$ values being higher than $D_N^{alloy/silicate}$ in the explored parameter space. It is important to note that even though the absolute proportions of C and N in cores and MOs are different for EMO and IMO differentiation regimes, the relative proportions of N and C in cores and MOs (controlled by $D_N^{alloy/silicate}$ and $D_C^{alloy/silicate}$ values) is similar for both EMOs and IMOs.

Due to the overwhelming partitioning of N and C into the atmospheres during EMO differentiation, our simulations show that EMO differentiation regime satisfies N and C contents of the parent cores for distinctly higher bulk N and C contents involved during core formation relative to IMO differentiation (Fig. 4, 5). N contents of the primitive cores can be explained via IMO differentiation if $6_{Q1:4}^{Q3:8}$ ppm of N (8, 6, and 4 are the median, third interquartile (Q3), and first interquartile values (Q1), respectively), and EMO differentiation if $418_{Q1:192}^{Q3:898}$ ppm of N was available to participate in core formation (Fig. 5B). Similarly, reproducing C contents of primitive cores via IMO differentiation requires lower bulk C contents ($124_{Q1:86}^{Q3:166}$ ppm) to be involved during core formation than EMO differentiation ($4549_{Q1:2425}^{Q3:8626}$ppm) (Fig. 5B). The gap between IMO and EMO differentiation regimes in terms of both bulk N and C contents (Fig. 5A) likely represents a solution space of protoplanetary differentiation that is a hybrid of end-member IMO-EMO differentiation regimes. Such a scenario would entail alloy melt-silicate melt equilibration in an IMO within a crust that was either permeable or had localized surficial magma ponds (Hirschmann et al., 2021). Unlike bulk N and C contents, the other four independent parameters, i.e., core-mantle mass ratio, S content of the alloy melt, $fO_2$ of core-mantle differentiation, and size of the parent body, yield solutions across the entire parameter space for both IMO and EMO differentiation regimes (Fig. 6). These four parameters only change the relative distribution between MOs and cores at a fixed bulk N and C content for both EMOs and IMOs. The overall solution space for IMO and EMO differentiation is controlled by the amount of N and C that is present in the protoplanet during its core formation.

## 5. Magma ocean differentiation in the parent bodies of iron meteorites – Internal or External?

To satisfy the N and C contents of the parent cores of magmatic iron meteorites, the interquartile ranges of bulk N and C involved during core formation need to be significantly higher for EMOs (N = 192-898 ppm and C = 2425-8626 ppm) than IMOs (N = 4-8 ppm and C = 86-166 ppm). N and C contents of several groups of volatile-depleted chondrites lie within the range of values required for EMO differentiation. Whereas, N and C contents required for IMO differentiation are sub-chondritic. Therefore, based on the chondritic abundances of N and C alone, both EMOs and IMOs present viable solutions for core formation in IMPBs. Chondrites, depending upon their metamorphic grade, sample surface to near-surface reservoirs of the protoplanets (Elkins-Tanton et al., 2011). Protoplanetary interiors experienced significantly higher temperatures than the surface layers during core formation (Neumann et al., 2012). The N and C inventories of the protoplanetary interiors during core formation may differ substantially from the chondritic abundances if N- and C-bearing hosts underwent significant alteration and decomposition before the onset of large-scale melting. A substantial drop in bulk N and C contents with increasing degree of thermal metamorphism and hydrothermal alteration has indeed been reported in CCs and OCs of varying petrologic grade (Alexander et al., 1998; Pearson et al., 2006).

Can a comparison of the N and C contents of chondritic samples as a function of their peak metamorphic temperatures provide insights into the N and C inventories of protoplanetary interiors at temperatures relevant for core formation? In Figure 7 we have plotted the literature data of N and C contents of bulk chondrites (Alexander et al., 2012; Pearson et al., 2006) against their effective temperatures ($T_{eff}$) based on organic thermometry (Cody et al., 2008). The bulk N and C contents of chondrites exhibit a sharp drop at $T_{eff}$ ~400-500 °C losing ~65-99% of their primitive N and C inventories. Therefore, primitive materials lost a substantial portion of their N and C inventory during low-$T$ thermal metamorphism. The extremely reduced Indarch (an enstatite chondrite (EC)) with the highest recorded $T_{eff}$ (~950 °C) contains a higher amount of N (~19% of the N inventory of the most volatile-rich chondrites) than ordinary chondrites and CVs recording a $T_{eff}$ of ~500 °C. But its C content follows the C-depletion trend traced by other groups of chondrites. It must be noted that ECs underwent thermal metamorphism under anomalously reduced conditions and a significant proportion of their N and C inventory resides in refractory phases like osbornite (TiN), nierite (Si₃N₄), sinoite (Si₂N₂O), graphite, and carbides (Grady et al., 1986). A substantially higher $f$O₂ of core-mantle differentiation in IMPBs (Righter et al., 2016) would not allow for the formation of these reduced phases, thereby suppressing the amount of N and C retained at larger extents of thermal metamorphism.

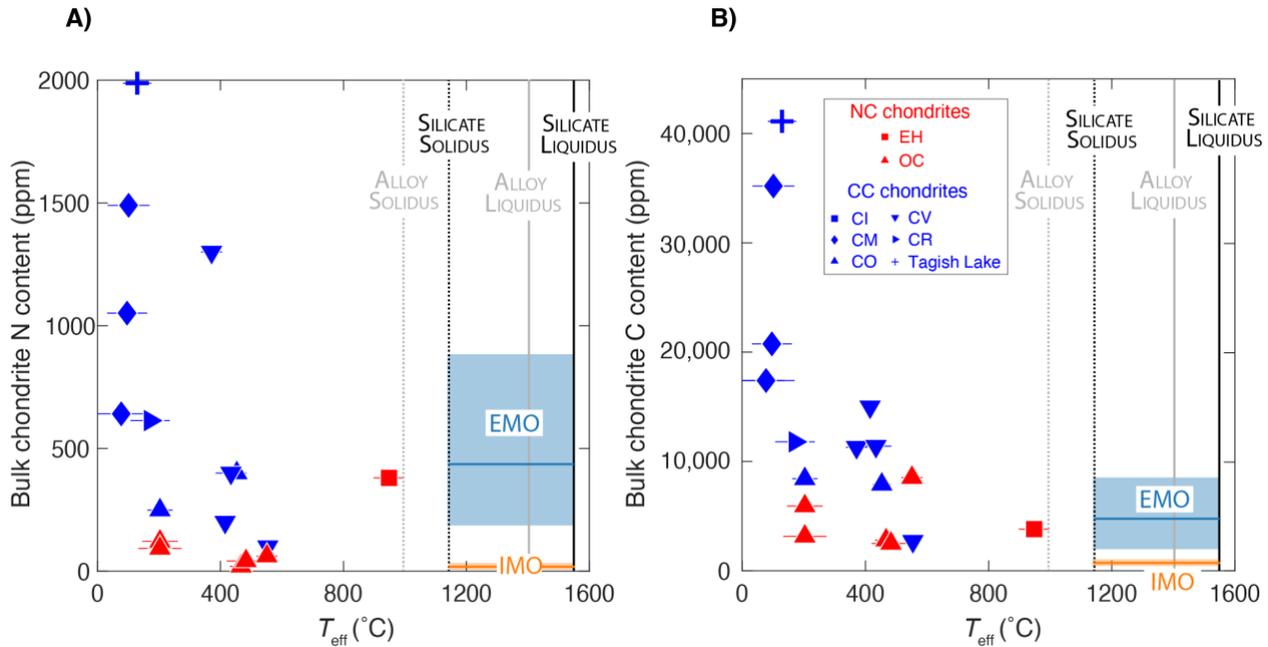

**Figure 7:** *Nitrogen and carbon contents in bulk chondrites as a function of effective temperatures (T$_{eff}$) of chondrites inferred from organic thermometry. Nitrogen and carbon contents in bulk chondrites decrease with increasing T$_{eff}$ suggesting loss of N and C with increasing degree of thermal metamorphism. Red and blue symbols represent chondrite groups belonging to the NC and CC cluster, respectively. Orange and blue lines represent the median values while the boxes represent the interquartile range of the solutions from inverse Monte Carlo simulations for IMOs and EMOs, respectively, within the temperature range of alloy melt-silicate melt equilibration. Data sources: N and C contents in bulk chondrites (Alexander et al., 2012; Pearson et al., 2006); T$_{eff}$ (Cody et al., 2008); alloy and silicate solidus and liquidus (Neumann et al., 2012).*

Based on higher temperatures during the onset of alloy and silicate melting relative to parent body metamorphism, protoplanetary interiors could have experienced more N and C loss than the surface layers (sampled by chondrites) during core formation (Fig. 7). It has been shown previously on the basis of permeability, porosity, and thermal state of protoplanetary interiors that positively buoyant N- and C-bearing fluids can percolate towards the surface of the protoplanets within hundreds of years (Hashizume and Sugiura, 1998; Sugiura et al., 1986). This period – much shorter than the time required for the formation of alloy and silicate melts (Kaminski et al., 2020) – can result in the efficient removal of the N- and C-bearing fluids from protoplanetary interiors before the onset of core formation. This would favor IMO differentiation in IMPBs because it allows for additional loss of N and C from their interiors at high temperatures relevant for core formation. However, it must be noted that inferring IMO differentiation as the dominant regime in IMPBs from the above evidence is contingent on several factors: 1) N and C inventories of protoplanetary interiors exhibiting a continuous drop with increasing temperatures (similar to the observations of chondrites in the low temperature regime ($T$ <~500 °C)). 2) The permeability of protoplanetary interiors is sufficient to allow for the positively buoyant N- and C-bearing fluids to freely percolate towards the surface. 3) The effect of pressure on N and C loss during thermal metamorphism should not be significant, i.e., the interior layers of a protoplanet, at a given temperature, must exhibit N and C loss comparable to chondritic samples even though they experienced thermal metamorphism at higher pressures. Whether some or all of these conditions are satisfied during thermal metamorphism of protoplanetary interiors is not well understood. Future work based on experiments and thermochemical modeling is necessary to test the validity of these factors. As a result, evidence for IMO differentiation based solely on N and C fractionation during protoplanetary differentiation is premature at this stage and cannot be used to definitively constrain the MO differentiation regime of the earliest formed protoplanets in the Solar System.

There are additional lines of evidence which present a more convincing case for the prevalence of IMOs during core formation in the earliest formed protoplanets. Accounting for the conductive and/or convective thermal evolution of the interiors of magmatic IMPBs, thermochemical models posit that molten silicates must be overlain by crusts composed of layers of undifferentiated, cold lids spanning a few kilometers (Hevey and Sanders, 2006; Kaminski et al., 2020; Sahijpal et al., 2007). The spread in chondrule ages (~0-3 Myrs after CAIs) of CV chondrites also suggests that the surfaces of protoplanets continued to accrete cold, primitive material as the interiors were undergoing differentiation (Elkins-Tanton et al., 2011). This continual accretion and a subsequent increase in the thickness of undifferentiated crusts (which act as a thermal boundary layer) further diminishes the possibility of the formation of EMOs. Additionally, the low gravitational force of asteroid-sized bodies makes it improbable for the earliest formed protoplanets to sustain MO degassed atmospheres for a period long enough to facilitate efficient atmosphere-MO-core equilibration (Young et al., 2019). Therefore, even if the these formed protoplanets had EMOs, a lack of vapor pressure build up should have caused a rapid loss of MO degassed HVEs to space resulting in the formation of almost N- and C-free cores and mantles (Grewal et al., 2021b). This is in disagreement with the available iron meteorite data which allude to a N- and C-bearing character of the parent cores of the earliest formed protoplanets.

Hirschmann et al. (2021) postulated EMO differentiation as a possible mechanism to explain the observed C/S ratios in the parent cores of magmatic iron meteorites because their IMO differentiation models predicted substantially higher amounts of C in the parent cores than what is observed in the iron meteorite data. The difference between the conclusions drawn in this study and those of Hirschmann et al. (2021) is tied to the C concentrations in the protoplanetary interiors during core formation. Hirschmann et al. (2021) estimated these values based on C abundances in different classes of chondrites ranging from least to most thermally altered (20,000, 8000 and 320 ppm C; 320 ppm value was based on the C content of LL6 chondrites). However, the peak temperatures experienced by even their most C-depleted end-member – LL6 chondrites – was only in the range of 900-960 °C (Huss et al., 2006). As core formation in protoplanetary interiors took place at significantly higher temperatures, even the most C-poor end-member of Hirschmann et al. (2021) could be an overestimate for the amount of C available during core formation.

Correspondingly, an involvement of lower than LL6 chondrite-derived C abundances (<165 ppm (Fig. 7B), i.e., lower than half of the lowest estimate used by Hirschmann et al. (2021), during core formation in IMPBs would result in no discrepancy between the findings of this study and those of Hirschmann et al. (2021).

## 6. Implications for volatile depletion in the earliest formed protoplanets

The presence of IMOs in IMPBs has important implications for the cause behind MVE depletion in other early forming protoplanets like Vesta and angrite parent body (APB). Vapor loss from the surface of an EMO has been postulated as a possible explanation for the depletion and associated isotope fractionation of MVEs like Cl and Cr in HEDs (Sarafian et al., 2017; Zhu et al., 2019). Lack of EMOs in contemporaneously accreting IMPBs means that alternate mechanisms are required to explain these observations. Accretion of MVE depleted precursors (Tian et al., 2019), impact induced-evaporation (O'Neill and Palme, 2008; Pringle et al., 2014), and surface volcanism (Abernethy et al., 2018) are other mechanisms postulated as explanations for the MVE depletion recorded by HEDs and angrites. Akin to EMO differentiation, the latter two processes also result in surficial vapor-silicate melt exchange albeit post MO differentiation and crystallization. As HEDs are basaltic rocks or cumulates, their MVE depletion can be biased towards the effect of these two processes. Therefore, if all protoplanets accreted materials with similar MVE abundances, a possible IMO differentiation regime in Vesta and APB suggests that localized processes like impact induced erosion and surface volcanism post-MO differentiation were the primary cause of MVE depletion recorded by basaltic rocks/cumulates like HEDs and angrites.

IMO differentiation in magmatic IMPBs suggests that metamorphic devolatilization followed by percolative flow of free fluids to the surface resulted in N- and C-poor interiors (relative to primitive chondritic materials) prior to the onset of core formation (Fig. 8). Equilibration of alloy and silicate melts in extremely N- and C-poor conditions resulted in N- and C-poor cores and mantles overlain by relatively N- and C-enriched undifferentiated crusts. This physical architecture resembles that of the parent bodies of several groups of chondrites like CV, CM, and Rumuriti where chondritic shells encased convective MOs and cores (e.g., Carporzen et al., 2011; Elkins-Tanton et al., 2011). The unmelted chondritic shells would comprise variably thermally and hydrothermally altered stratigraphy analogous to those sampled by carbonaceous, ordinary, and enstatite chondrites. The prevalence of IMOs with a stratified physical architecture in the earliest formed protoplanets implies that some groups of iron meteorites and chondrites were potentially genetically linked, i.e., they originated from same parent bodies. This has indeed been suggested for IIE irons-H chondrites and IVA irons-L/LL chondrites based on the similarity of their oxygen isotopic compositions and textural and mineralogical features (Goldstein et al., 2009).

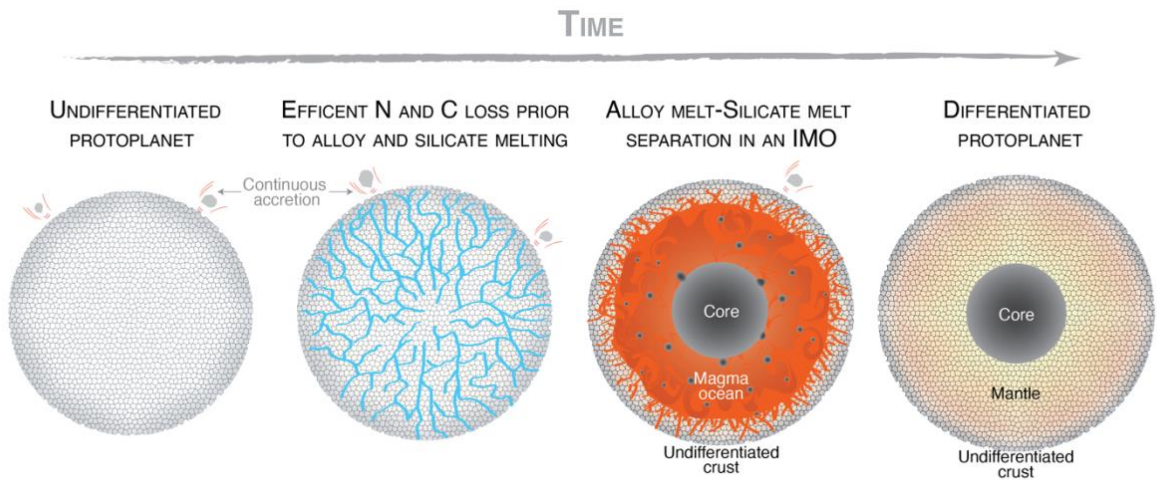

**Figure 8:** *An illustration showing the temporal evolution of the parent bodies of iron meteorites undergoing IMO differentiation. During progressive heating of a parent body, N and C are extensively lost from its interior prior to the onset of alloy and silicate melting. Alloy melts equilibrate with silicate melts followed by migration towards the center of the protoplanet. The MO is overlain by an undifferentiated crust, which is sintered and metamorphosed in its deeper layers and unsintered and primitive in its surface layers. Also, the surface may continue to accrete primitive material onto its surface increasing the crustal thickness. Post-differentiation, the protoplanet would compose of a differentiated mantle that is overlain by an undifferentiated crust and underlain by a metallic core.*

Akin to magmatic IMPBs, if IMO differentiation was the dominant differentiation mechanism in the earliest formed protoplanets across the Solar System, then the formation of differentiated cores and mantles overlain by solid undifferentiated shells must have been the norm. Metamorphic devolatilization, rather than irreversible sublimation of organics in the solar nebula (Li et al., 2021) and loss of MO degassed atmospheres from EMOs (Grewal et al., 2021b; Hirschmann et al., 2021), was likely the primary cause of N and C depletion within protoplanetary interiors. Water loss in protoplanets accreting at various heliocentric distances as well as C loss from protoplanets that specifically accreted CO ice-bearing dust (>20 AU) has also been explained by similar devolatilization processes powered by radiogenic heating (Lichtenberg et al., 2019; Lichtenberg and Krijt, 2021). Siderophile HVEs like N and C predominantly resided either in the metallic cores of differentiated interiors or within undifferentiated surficial layers. N and C isotopic compositions of the cores of inner and outer Solar System protoplanets could also be imprinted by the effects of parent body processing on N- and C- bearing primitive materials. Crustal layers, depending upon the extent of parent body processing, could retain the isotopic compositions of primitive materials.

## 7. Concluding remarks

In this study, we used two end-member core formation scenarios – IMO and EMO – to explain the N and C abundances in the parent cores of magmatic IMPBs. EMO differentiation necessitates the involvement of ~10-45% and ~5-17% of chondritic N and C inventories, respectively, during core formation to explain N and C contents in the parent cores of these protoplanets. Whereas, IMO differentiation requires the involvement of ~0.2-0.4% and ~0.2-0.3% of chondritic N and C inventories. N and C contents required for EMO differentiation lie within the range of several groups of volatile-depleted chondrites. Whereas N and C contents required for IMO differentiation are sub-chondritic. Evidence from the chondrite record suggests that radiogenic heating led to significant parent body processing in protoplanetary interiors prior to the onset of core formation. Provided substantial loss of primitive N and C inventories before the formation of alloy and silicate melts, N and C contents of the parent cores of magmatic irons can be satisfied more easily by IMO differentiation. This suggests that the earliest formed protoplanets across the Solar System consisted of differentiated mantles overlain by undifferentiated crusts (which were primitive, unsintered on the surface and metamorphosed, sintered in the deeper layers) and underlain by metallic cores. Also, the prevalence of IMO differentiation regime in the earliest formed protoplanets suggests that metamorphic devolatilization prior to the formation of alloy and silicate melts was the primary cause of N and C depletion in the differentiated rocky bodies of the Solar System. The findings of this study hinge on two important assumptions - 1) $D_N^{alloy/silicate}$ and $D_C^{alloy/silicate}$ determined for systems containing wt.% N and C are applicable to natural systems containing ppm level N and C. 2) N- and C-loss as a function of thermal metamorphism observed in surficial layers (as sampled by chondrites) is also applicable to protoplanetary interiors at higher pressures. Future experimental and numerical work on these fronts would test the validity of these assumptions.

## Acknowledgements

Amrita P. Vyas created the final versions of the schematic diagrams in Figures 1 and 8. D.S.G. thanks Linda Elkins-Tanton for providing the seed schematic diagrams for Figures 1 and 8. Comments by Fabrice Gaillard and Colin Jackson as reviewers of this manuscript are gratefully acknowledged. Paolo Sossi and four anonymous reviewers are also thanked for their comments on earlier versions of this manuscript, which significantly improved our communication. D.S.G. received support from a NASA FINESST grant 80NSSC19K1538, a Lodieska Stockbridge Vaughn Fellowship by Rice University, and a Barr Foundation Postdoctoral Fellowship by California Institute of Technology. R.D. received support from NASA grants 80NSSC18K0828 and 80NSSC18K1314.

## References:

Abernethy, F.A.J., Verchovsky, A.B., Franchi, I.A., Grady, M.M., 2018. Basaltic volcanism on the angrite parent body: Comparison with 4 Vesta. Meteoritics & Planetary Science 53, 375–393. https://doi.org/10.1111/maps.13016

Abernethy, F.A.J., Verchovsky, A.B., Starkey, N.A., Anand, M., Franchi, I.A., Grady, M.M., 2013. Stable isotope analysis of carbon and nitrogen in angrites. Meteoritics & Planetary Science 48, 1590–1606. https://doi.org/10.1111/maps.12184

Alexander, C.M.O., Bowden, R., Fogel, M.L., Howard, K.T., Herd, C.D.K., Nittler, L.R., 2012. The Provenances of Asteroids, and Their Contributions to the Volatile Inventories of the Terrestrial Planets. Science (1979) 337, 721–723. https://doi.org/10.1126/science.1223474

Alexander, C.M.O., Russell, S.S., Arden, J.W., Ash, R.D., Grady, M.M., Pillinger, C.T., 1998. The origin of chondritic macromolecular organic matter: A carbon and nitrogen isotope study. Meteoritics & Planetary Science 33, 603–622. https://doi.org/10.1111/j.1945-5100.1998.tb01667.x

Alexander, C.M.O.D., Fogel, M., Yabuta, H., Cody, G.D., 2007. The origin and evolution of chondrites recorded in the elemental and isotopic compositions of their macromolecular organic matter. Geochimica et Cosmochimica Acta 71, 4380–4403. https://doi.org/10.1016/j.gca.2007.06.052

Carporzen, L., Weiss, B.P., Elkins-Tanton, L.T., Shuster, D.L., Gattacceca, J., 2011. Magnetic evidence for a partially differentiated carbonaceous chondrite parent body. Proc Natl Acad Sci U S A 108, 6386–6389. https://doi.org/10.1073/pnas.1017165108

Chabot, N.L., 2004. Sulfur contents of the parental metallic cores of magmatic iron meteorites. Geochimica et Cosmochimica Acta 68, 3607–3618. https://doi.org/10.1016/j.gca.2004.03.023

Cody, G.D., Alexander, C.M.O., Yabuta, H., Kilcoyne, A.L.D., Araki, T., Ade, H., Dera, P., Fogel, M., Militzer, B., Mysen, B.O., 2008. Organic thermometry of the chondritic parent bodies. Earth and Planetary Science Letters 272, 446–455. https://doi.org/10.1016/j.epsl.2008.05.008

Dasgupta, R., Chi, H., Shimizu, N., Buono, A.S., Walker, D., 2013. Carbon solution and partitioning between metallic and silicate melts in a shallow magma ocean: Implications for the origin and distribution of terrestrial carbon. Geochimica et Cosmochimica Acta 102, 191–212. https://doi.org/10.1016/j.gca.2012.10.011

Dasgupta, R., Grewal, D.S., 2019. Origin and Early Differentiation of Carbon and Associated Life-Essential Volatile Elements on Earth, in: Orcutt, B., Daniel, I., Dasgupta, R. (Eds.), Deep Carbon: Past to Present. Cambridge University Press, pp. 4–39. https://doi.org/10.1017/9781108677950.002

Dasgupta, R., Walker, D., 2008. Carbon solubility in core melts in a shallow magma ocean environment and distribution of carbon between the Earth's core and the mantle. Geochimica et Cosmochimica Acta 72, 4627–4641. https://doi.org/10.1016/j.gca.2008.06.023

Elkins-Tanton, L.T., Weiss, B.P., Zuber, M.T., 2011. Chondrites as samples of differentiated planetesimals. Earth and Planetary Science Letters 305, 1–10. https://doi.org/10.1016/j.epsl.2011.03.010

Fischer, R.A., Cottrell, E., Hauri, E., Lee, K.K.M., le Voyer, M., 2020. The carbon content of Earth and its core. Proceedings of the National Academy of Sciences 201919930. https://doi.org/10.1073/pnas.1919930117

Gaillard, F., Malavergne, V., Bouhifd, M.A., Rogerie, G., 2022. A speciation model linking the fate of carbon and hydrogen during core-magma ocean equilibration. Earth and Planetary Science Letters 577, 117266. https://doi.org/10.1016/j.epsl.2021.117266

Goldstein, J.I., Huss, G.R., Scott, E.R.D., 2017. Ion microprobe analyses of carbon in Fe–Ni metal in iron meteorites and mesosiderites. Geochimica et Cosmochimica Acta 200, 367–407. https://doi.org/10.1016/j.gca.2016.12.027

Goldstein, J.I., Scott, E.R.D., Chabot, N.L., 2009. Iron meteorites: Crystallization, thermal history, parent bodies, and origin. Chemie der Erde 69, 293–325. https://doi.org/10.1016/j.chemer.2009.01.002

Grady, M.M., Wright, I.P., Carr, L.P., Pillinger, C.T., 1986. Compositional differences in enstatite chondrites based on carbon and nitrogen stable isotope measurements. Geochimica et Cosmochimica Acta 50, 2799–2813. https://doi.org/10.1016/0016-7037(86)90228-0

Grewal, D.S., Dasgupta, R., Aithala, S., 2021a. The effect of carbon concentration on its core-mantle partitioning behavior in inner Solar System rocky bodies. Earth and Planetary Science Letters 571, 117090. https://doi.org/10.1016/j.epsl.2021.117090

Grewal, D.S., Dasgupta, R., Farnell, A., 2020. The speciation of carbon, nitrogen, and water in magma oceans and its effect on volatile partitioning between major reservoirs of the Solar System rocky bodies. Geochimica et Cosmochimica Acta 280, 281–301. https://doi.org/10.1016/j.gca.2020.04.023

Grewal, D.S., Dasgupta, R., Holmes, A.K., Costin, G., Li, Y., Tsuno, K., 2019a. The fate of nitrogen during core-mantle separation on Earth. Geochimica et Cosmochimica Acta 251, 87–115. https://doi.org/10.1016/j.gca.2019.02.009

Grewal, D.S., Dasgupta, R., Hough, T., Farnell, A., 2021b. Rates of protoplanetary accretion and differentiation set nitrogen budget of rocky planets. Nature Geoscience 14, 369–376. https://doi.org/10.1038/s41561-021-00733-0

Grewal, D.S., Dasgupta, R., Marty, B., 2021c. A very early origin of isotopically distinct nitrogen in inner Solar System protoplanets. Nature Astronomy 5, 356–364. https://doi.org/10.1038/s41550-020-01283-y

Grewal, D.S., Dasgupta, R., Sun, C., Tsuno, K., Costin, G., 2019b. Delivery of carbon, nitrogen, and sulfur to the silicate Earth by a giant impact. Science Advances 5, eaau3669. https://doi.org/10.1126/sciadv.aau3669

Hashizume, K., Sugiura, N., 1998. Transportation of gaseous elements and isotopes in a thermally evolving chondritic planetesimal. Meteoritics & Planetary Science 33, 1181–1195. https://doi.org/10.1111/j.1945-5100.1998.tb01722.x

Hevey, P.J., Sanders, I.S., 2006. A model for planetesimal meltdown by 26 Al and its implications for meteorite parent bodies. Meteoritics & Planetary Science 41, 95–106. https://doi.org/10.1111/j.1945-5100.2006.tb00195.x

Hin, R.C., Coath, C.D., Carter, P.J., Nimmo, F., Lai, Y.J., Pogge von Strandmann, P.A.E., Willbold, M., Leinhardt, Z.M., Walter, M.J., Elliott, T., 2017. Magnesium isotope evidence that accretional vapour loss shapes planetary compositions. Nature 549, 511–527. https://doi.org/10.1038/nature23899

Hirschmann, M.M., Bergin, E.A., Blake, G.A., Ciesla, F.J., Li, J., 2021. Early volatile depletion on planetesimals inferred from C–S systematics of iron meteorite parent bodies. Proceedings of the National Academy of Sciences 118, e2026779118. https://doi.org/10.1073/pnas.2026779118

Huss, G.R., Rubin, A.E., Grossman, J.N., 2006. Thermal Metamorphism in Chondrites, in: Lauretta, D.S., McSween Jr., H.Y. (Eds.), Meteorites and the Early Solar System II. University of Arizona Press, pp. 567–586.

Kaminski, E., Limare, A., Kenda, B., Chaussidon, M., 2020. Early accretion of planetesimals unraveled by the thermal evolution of the parent bodies of magmatic iron meteorites. Earth and Planetary Science Letters 548, 116469. https://doi.org/10.1016/j.epsl.2020.116469

Keppler, H., Golabek, G., 2019. Graphite floatation on a magma ocean and the fate of carbon during core formation. Geochemical Perspectives Letters 11, 12–17. https://doi.org/10.7185/geochemlet.1918

Kruijer, T.S., Burkhardt, C., Budde, G., Kleine, T., 2017. Age of Jupiter inferred from the distinct genetics and formation times of meteorites. Proceedings of the National Academy of Sciences 201704461. https://doi.org/10.1073/pnas.1704461114

Kruijer, T.S., Touboul, M., Fischer-Godde, M., Bermingham, K.R., Walker, R.J., Kleine, T., 2014. Protracted core formation and rapid accretion of protoplanets. Science (1979) 344, 1150–1154. https://doi.org/10.1126/science.1251766

Kuwahara, H., Itoh, S., Nakada, R., Irifune, T., 2019. The Effects of Carbon Concentration and Silicate Composition on the Metal-Silicate Partitioning of Carbon in a Shallow Magma Ocean. Geophysical Research Letters 46, 9422–9429. https://doi.org/10.1029/2019GL084254

Kuwahara, H., Itoh, S., Suzumura, A., Nakada, R., Irifune, T., 2021. Nearly carbon-saturated magma oceans in planetary embryos during core formation. Geophysical Research Letters 1–25. https://doi.org/10.1029/2021GL092389

Lewis, C.F., Moore, C.B., 1971. Chemical Analyses of Thirty-Eight Iron Meteorites. Meteoritics 6, 195–205. https://doi.org/10.1111/j.1945-5100.1971.tb00111.x

Lewis, J.S., Barshay, S.S., Noyes, B., 1979. Primodial retention of carbon by the terrestrial planets. Icarus 37, 190–206. https://doi.org/10.1016/0019-1035(79)90125-8

Li, J., Bergin, E.A., Blake, G.A., Ciesla, F.J., Hirschmann, M.M., 2021. Earth's carbon deficit caused by early loss through irreversible sublimation. Science Advances 7, 3–8. https://doi.org/10.1126/sciadv.abd3632

Libourel, G., Marty, B., Humbert, F., 2003. Nitrogen solubility in basaltic melt. Part I. Effect of oxygen fugacity. Geochimica et Cosmochimica Acta 67, 4123–4135. https://doi.org/10.1016/S0016-7037(03)00259-X

Lichtenberg, T., Golabek, G.J., Burn, R., Meyer, M.R., Alibert, Y., Gerya, T. v, Mordasini, C., 2019. A water budget dichotomy of rocky protoplanets from 26 Al-heating. Nature Astronomy. https://doi.org/10.1038/s41550-018-0688-5

Lichtenberg, T., Krijt, S., 2021. System-level Fractionation of Carbon from Disk and Planetesimal Processing. The Astrophysical Journal Letters 913, L20. https://doi.org/10.3847/2041-8213/abfdce

Maurel, C., Bryson, J.F.J., Lyons, R.J., Ball, M.R., Chopdekar, R. v, Scholl, A., Ciesla, F.J., Bottke, W.F., Weiss, B.P., 2020. Meteorite evidence for partial differentiation and protracted accretion of planetesimals. Science Advances 6, eaba1303. https://doi.org/10.1126/sciadv.aba1303

McSween, H.Y., 1989. Achondrites and igneous processes on asteroids. Annual review of earth and planetary sciences. Vol. 17 17, 119–140. https://doi.org/10.1146/annurev.ea.17.050189.001003

Neumann, W., Breuer, D., Spohn, T., 2012. Differentiation and core formation in accreting planetesimals. Astronomy and Astrophysics 543. https://doi.org/10.1051/0004-6361/201219157

Ni, H., Keppler, H., 2013. Carbon in Silicate Melts. Reviews in Mineralogy and Geochemistry 75, 251–287. https://doi.org/10.2138/rmg.2013.75.9

O'Neill, H.S.C., Palme, H., 2008. Collisional erosion and the non-chondritic composition of the terrestrial planets. Philosophical Transactions of the Royal Society A: Mathematical, Physical and Engineering Sciences 366, 4205–4238. https://doi.org/10.1098/rsta.2008.0111

Palme, H., Lodders, K., Jones, A., 2014. Solar System Abundances of the Elements, in: Treatise on Geochemistry. pp. 15–36. https://doi.org/10.1016/B978-0-08-095975-7.00118-2

Pearson, V.K., Sephton, M.A., Franchi, I.A., Gibson, J.M., Gilmour, I., 2006. Carbon and nitrogen in carbonaceous chondrites: Elemental abundances and stable isotopic compisitions. Meteoritics and Planetary Science 41, 1899–1918. https://doi.org/10.1111/j.1945-5100.2006.tb00459.x

Pringle, E.A., Moynier, F., Savage, P.S., Badro, J., Barrat, J.-A., 2014. Silicon isotopes in angrites and volatile loss in planetesimals. Proceedings of the National Academy of Sciences 111, 17029–17032. https://doi.org/10.1073/pnas.1418889111

Prombo, C.A., Clayton, R.N., 1993. Nitrogen isotopic compositions of iron meteorites. Geochimica et Cosmochimica Acta 57, 3749–3761. https://doi.org/10.1016/0016-7037(93)90153-N

Righter, K., Sutton, S.R., Danielson, L., Pando, K., Newville, M., 2016. Redox variations in the inner solar system with new constraints from vanadium XANES in spinels. American Mineralogist. https://doi.org/10.2138/am-2016-5638

Sahijpal, S., Soni, P., Gupta, G., 2007. Numerical simulations of the differentiation of accreting planetesimals with 26Al and 60Fe as the heat sources. Meteoritics and Planetary Science 42, 1529–1548. https://doi.org/10.1111/j.1945-5100.2007.tb00589.x

Sarafian, A.R., John, T., Roszjar, J., Whitehouse, M.J., 2017. Chlorine and hydrogen degassing in Vesta's magma ocean. Earth and Planetary Science Letters. https://doi.org/10.1016/j.epsl.2016.10.029

Sephton, M.A., Verchovsky, A.B., Bland, P.A., Gilmour, I., Grady, M.M., Wright, I.P., 2003. Investigating the variations in carbon and nitrogen isotopes in carbonaceous chondrites. Geochimica et Cosmochimica Acta 67, 2093–2108. https://doi.org/10.1016/S0016-7037(02)01320-0

Speelmanns, I.M., Schmidt, M.W., Liebske, C., 2018. Nitrogen Solubility in Core Materials. Geophysical Research Letters 45, 7434–7443. https://doi.org/10.1029/2018GL079130

Steenstra, E.S., Knibbe, J.S., Rai, N., van Westrenen, W., 2016. Constraints on core formation in Vesta from metal–silicate partitioning of siderophile elements. Geochimica et Cosmochimica Acta 177, 48–61. https://doi.org/10.1016/j.gca.2016.01.002

Steenstra, E.S., Sitabi, A.B., Lin, Y.H., Rai, N., Knibbe, J.S., Berndt, J., Matveev, S., van Westrenen, W., 2017. The effect of melt composition on metal-silicate partitioning of siderophile elements and constraints on core formation in the angrite parent body. Geochimica et Cosmochimica Acta 212, 62–83. https://doi.org/10.1016/j.gca.2017.05.034

Sugiura, N., Arkani-Hamed, J., Strangway, D.W., 1986. Possible transport of carbon in meteorite parent bodies, Earth and Planetary Science Letters.

Sugiura, N., Fujiya, W., 2014. Correlated accretion ages and ε 54 Cr of meteorite parent bodies and the evolution of the solar nebula. Meteoritics & Planetary Science 49, 772–787. https://doi.org/10.1111/maps.12292

Taylor, G.J., 1992. Core formation in asteroids. Journal of Geophysical Research 97, 14717. https://doi.org/10.1029/92JE01501

Tian, Z., Chen, H., Fegley, B., Lodders, K., Barrat, J.A., Day, J.M.D., Wang, K., 2019. Potassium isotopic compositions of howardite-eucrite-diogenite meteorites. Geochimica et Cosmochimica Acta 266, 611–632. https://doi.org/10.1016/j.gca.2019.08.012

Tsuno, K., Grewal, D.S., Dasgupta, R., 2018. Core-mantle fractionation of carbon in Earth and Mars: The effects of sulfur. Geochimica et Cosmochimica Acta 238, 477–495. https://doi.org/10.1016/j.gca.2018.07.010

Worsham, E.A., Burkhardt, C., Budde, G., Fischer-Gödde, M., Kruijer, T.S., Kleine, T., 2019. Distinct evolution of the carbonaceous and non-carbonaceous reservoirs: Insights from Ru, Mo, and W isotopes. Earth and Planetary Science Letters 521, 103–112. https://doi.org/10.1016/j.epsl.2019.06.001

Yang, J., Goldstein, J.I., Michael, J.R., Kotula, P.G., Scott, E.R.D., 2010. Thermal history and origin of the IVB iron meteorites and their parent body. Geochimica et Cosmochimica Acta 74, 4493–4506. https://doi.org/10.1016/j.gca.2010.04.011

Yang, J., Goldstein, J.I., Scott, E.R.D., 2008. Metallographic cooling rates and origin of IVA iron meteorites. Geochimica et Cosmochimica Acta 72, 3043–3061. https://doi.org/10.1016/j.gca.2008.04.009

Yoshioka, T., Nakashima, D., Nakamura, T., Shcheka, S., Keppler, H., 2019. Carbon solubility in silicate melts in equilibrium with a CO-CO2 gas phase and graphite. Geochimica et Cosmochimica Acta. https://doi.org/10.1016/j.gca.2019.06.007

Young, E.D., Shahar, A., Nimmo, F., Schlichting, H.E., Schauble, E.A., Tang, H., Labidi, J., 2019. Near-equilibrium isotope fractionation during planetesimal evaporation. Icarus 323, 1–15. https://doi.org/10.1016/j.icarus.2019.01.012

Zhu, K., Sossi, P.A., Siebert, J., Moynier, F., 2019. Tracking the volatile and magmatic history of Vesta from chromium stable isotope variations in eucrite and diogenite meteorites. Geochimica et Cosmochimica Acta. https://doi.org/10.1016/j.gca.2019.07.043